\documentclass{svmult} % We use the Springer-Verlag format

\usepackage{graphicx}        % standard LaTeX graphics tool
\usepackage[bottom]{footmisc}% places footnotes at page bottom

\usepackage{amssymb}

\newtheorem{theor}{Theorem}[section]
\newtheorem{defin}{Definition}[section]

\newtheorem{corol}{Corollary}[section]

\begin{document}

\title*{A simple proof of the Jamio\l kowski criterion for complete positivity of linear maps of algebras of Hilbert-Schmidt operators}
%\footnote{This paper is dedicated to Prof.\ Alberto Galindo on the occasion of his 70th birthday.}
\titlerunning{A simple proof of the Jamio\l kowski criterion}
\author{David Salgado\and
Jos\'{e} L.\ S\'{a}nchez-G\'{o}mez%\and
%Miguel Ferrero\inst{2}
}
\institute{Dpto.\ F\'{\i}sica Te\'{o}rica, Univ.\ Aut\'{o}noma de Madrid\\
\texttt{david.salgado@uam.es; jl.sanchezgomez@uam.es}
%\and Dpto.\ F\'{\i}sica, Univ.\ de Oviedo \texttt{maferrero@uniovi.es}
}

\maketitle

\footnote{This paper is dedicated to Prof.\ Alberto Galindo on the occasion of his 70th birthday.}We generalize a preceding simple proof of the Jamio\l kowski criterion to check whether a given linear map between algebras of operators is completely positive or not. The generalization is performed to embrace all algebras of Hilbert-Schmidt class operators, thus possibly infinite-dimensional.
%Your text goes here. Separate text sections with the standard \LaTeX\ sectioning commands.

\section{Introduction}
\label{Intro}
% Always give a unique label
% and use \ref{<label>} for cross-references
% and \cite{<label>} for bibliographic references
% use \sectionmark{}
% to alter or adjust the section heading in the running head
%Your text goes here. Use the \LaTeX\ automatism for your citations \cite{monograph}.
Complete positivity \cite{Sti55a,Sto63a,Arv69a,Cho82a} is a mathematical property of some maps between $C^{*}$ algebras, typically algebras of operators, which arises in Quantum Physics in two relevant fields, namely the dynamics of open quantum systems \cite{Dav76a,Lin76a,GorKosSud76a,AliLen87a,BrePet02a,BucHor02a} and the analysis of entanglement and separability of quantum states \cite{Per96a,HorHorHor96a,Lewetal00a,Brus01a}.\\
In this contribution we generalize a simple proof of the so-called Jamio\l kowski criterion \cite{dePil67a,Jam72a} to chech whether a given linear map is completely positive (CP herafter) or not. Our preceding proof \cite{SalSanFer04a} (see also \cite{Sal04a}) only applies to maps between finite algebras, i.e.\ to finite quantum systems. Here we extend the proof to include also any algebra of Hilbert-Schmidt operators.\\
The work is divided as follows. In section \ref{JamCri} we enunciate the Jamio\l kowski criterion and present the main tools and preliminary results to be used later in section \ref{Pro}, where the proof is worked out in detail. In section \ref{Exa} we include examples to illustrate the usage of these results. In section \ref{DisCon} we end with a brief discussion and some conclusions.

\section{Jamio\l kowski criterion}
\label{JamCri}

The rigourous definition of complete positivity is 

\begin{defin}
Let $\mathfrak{A}$ and $\mathfrak{B}$ be two $C^{*}$ algebras. Let $\alpha$ be a linear map $\alpha:\mathfrak{A}\to\mathfrak{B}$. The linear map $\alpha$ is CP if the extended map $\alpha\otimes\mathbb{I}_{N}:\mathfrak{A}\otimes\mathcal{M}_{N}(\mathbb{C})\to\mathfrak{B}\otimes\mathcal{M}_{N}(\mathbb{C})$, where $\mathcal{M}_{N}(\mathbb{C})$ denotes the algebra of complex matrices of dimension $N$ and $\mathbb{I}_{N}$ the identity map on this algebra, is positive for all $N\geq 1$.
\end{defin}

This definition is scarcely used in practical applications and more manageable criteria are desired. In this respect, in quantum physics the most relevant characterization of CP maps is known as the Kraus decomposition \cite{Kra71a,Cho75a}, which we enunciate here only for the case of endomorphisms, since it shows more relevance for physical applications:

\begin{theor}{(Kraus representation)}
A given linear endomorphism $\alpha$ on a $C^{*}$ algebra $\mathfrak{A}$ is CP if, and only if, it can be written as 

\begin{equation}\label{KraRep}
\alpha[X]=\sum_{k\in\mathcal{I}}M_{k}XM_{k}^{\dagger}\qquad\forall X\in\mathfrak{A}
\end{equation}

\noindent where the set of elements $\{M_{k}\}_{k\in\mathcal{I}}$ (possibly infinite) is arbitrary\footnote{Up to further requirements upon $\alpha$.}.
\end{theor}

The set of the so-called Kraus operators $M_{k}$ is not unique. In the case of countable infinite Kraus operators, the series is understood to converge in the norm topology.\\

Provided one can find a valid set of Kraus operators, the complete positivity of a given linear map is assured, but to find these operators is not immediate in practice. Thus a simpler criterion is needed. In this respect, the Jamio\l kowski criterion provides a systematic way to know whether a linear map is CP or not, but the original proof did not either provide a method to find such operators; the following does. The Jamio\l kowski criterion for finite complex matrices algebras reads

\begin{theor}{(Jamio\l kowski criterion)}
Let $\alpha$ be a linear endomorphism upon $\mathcal{M}_{N}(\mathbb{C})$. Let $\mathcal{J}_{e}[\alpha]$ be the element in $\mathcal{M}_{N}(\mathbb{C})\otimes\mathcal{M}_{N}(\mathbb{C})$ given by  

\begin{equation}
\mathcal{J}_{e}[\alpha]=\sum_{ij=1}^{N}\alpha(E_{ij})\otimes E_{ij}
\end{equation}

\noindent where $E_{ij}$ denotes a generalized Weyl basis of $\mathcal{M}_{N}(\mathbb{C})$ in the orthonormal basis $\{e_{k}\}_{k=1,\dots,N}$ of $\mathbb{C}^{N}$. Then $\alpha$ is CP if, and only if, $\mathcal{J}_{e}[\alpha]\geq 0$.  

\end{theor}

Our proof of this result \cite{SalSanFer04a} allows us to find \emph{any} set of Kraus matrices of a CP linear map as well as the number of Kraus matrices of a minimal Kraus representation. To extend this result to (possibly infinite-dimensional) algebras of Hilbert-Schmidt operators, we must make clear the following result\footnote{Nearly all the mathematical definitions, properties and theorems involved in the subsequent proof can be found in \cite{AbeGal91a}, from which one of the present authors had  his first contact with the widely used techniques of Hilbert spaces.}: the space $\mathcal{C}_{2}(\mathfrak{H})$ of Hilbert-Schmidt class operators upon a complex separable Hilbert space $\mathfrak{H}$ admits a Hilbert space structure by defining the inner product in it by $$(A,B)_{2}\equiv\textrm{tr}\left(A^{\dagger}B\right)$$where $A$ and $B$ are Hilbert-Schmidt operators (Proposition 9.12 of \cite{AbeGal91a}). Thus all Hilbert-space techniques can be used, in particular, after convincing oneself that $\{P_{ij}\}_{i,j\geq 1}$ is an orthonormal Hilbert basis\footnote{Just use Parseval's identity $(A,B)_{2}=\sum_{ij=1}^{\infty}(A,P_{ij})_{2}(P_{ij},B)_{2}$ for all $A,B\in\mathcal{C}_{2}(\mathfrak{H})$.}, where $P_{ij}$ denotes, in common Physics notation, $P_{ij}\equiv|e_{i}\rangle\langle e_{j}|$ ($\{e_{k}\}$ denoting a Hilbert basis of $\mathfrak{H}$), one can write for an arbitrary element $X$ of $\mathcal{C}_{2}$:

\begin{equation}\label{FouExp}
X=\sum_{ij=1}^{\infty}(P_{ij},X)_{2}P_{ij}=\sum_{ij=1}^{\infty}(e_{j},Xe_{i})_{\mathfrak{H}}P_{ij}
\end{equation}

This, in conjuction with the following characterization of positive elements of a $C^{*}$-algebra, sets the tools for the forthcoming proof: An element $X$ of a $C^{*}$-algebra $\mathcal{A}$ is positive if, and only if, it can be written as $X=B^{*}B$, where $B$ is an arbitrary element of $\mathfrak{A}$ (cf.\ \cite{BraRob87a} for details).

\section{The proof}
\label{Pro}

The extension of the previous Jamio\l kowski criterion reads

\begin{theor}{(Jamio\l kowski criterion)}
Let $\alpha$ be a linear endomorphism on the space of Hilbert-Schmidt operators $\mathcal{C}_{2}(\mathfrak{H})$ upon a complex separable Hilbert space $\mathfrak{H}$. Let $\mathcal{J}_{e}[\alpha]$ be the element of $\mathcal{C}_{2}(\mathfrak{H})\otimes\mathcal{C}_{2}(\mathfrak{H})$ given by

\begin{equation}
\mathcal{J}_{e}[\alpha]=\sum_{ij=1}^{\infty}\alpha[P_{ij}]\otimes P_{ij}
\end{equation}

\noindent where $P_{ij}\equiv|e_{i}\rangle\langle e_{j}|$, as before. Then $\alpha$ is CP if, and only if, $\mathcal{J}_{e}[\alpha]\geq 0$.

\end{theor}

\begin{proof}
\noindent $(\Rightarrow)$ First we will prove that if $\alpha$ is CP, then $\mathcal{J}_{e}[\alpha]$ is a positive element. Let us begin by applying the Fourier expansion (\ref{FouExp}) to the elements $\alpha[P_{ij}]$ themselves:

\begin{equation}\label{LinPij}
\alpha[P_{ij}]=\sum_{nm=1}^{\infty}\alpha_{ijmn}P_{mn}
\end{equation}

\noindent where $\alpha_{ijmn}=(P_{mn},\alpha[P_{ij}])_{2}$. If $\alpha$ is CP, then there exists a set of Kraus operators such that 

%\begin{subequation}
\begin{equation}
\alpha[X]=\sum_{p=1}^{\infty}M_{p}XM_{p}^{\dagger}
\end{equation}
\noindent and, in particular, 

\begin{equation}\label{KraPij}
\alpha[P_{ij}]=\sum_{p=1}^{\infty}M_{p}P_{ij}M_{p}^{\dagger}
\end{equation}
%\end{subequation}
\noindent Using (\ref{LinPij}) together with (\ref{KraPij}) as well as general properties of inner products and adjoint operators, one can immediately find 

\begin{equation}
\alpha_{ijmn}=\sum_{p=1}^{\infty}(e_{n},M_{p}e_{j})_{\mathfrak{H}}^{*}(e_{m},M_{p}e_{i})_{\mathfrak{H}}
\end{equation}

We then define the sequences $f_{ij}\equiv\left((e_{i},M_{1}e_{j})_{\mathfrak{H}},(e_{i},M_{2}e_{j})_{\mathfrak{H}},(e_{i},M_{3}e_{j})_{\mathfrak{H}},\dots\right)$ It is easy to convince oneself, given that $M_{p}\in\mathcal{C}_{2}(\mathfrak{H})$ for all $p$, that $f_{ij}\in\ell^{2}(\mathbb{C})$. Thus we have

\begin{equation}
\alpha_{ijmn}=(f_{nj},f_{mi})_{\ell^{2}(\mathbb{C})}
\end{equation}

The question is then reduced to check whether

\begin{equation}
\sum_{ij=1}^{\infty}\sum_{mn=1}^{\infty}(f_{nj},f_{mi})_{\ell^{2}(\mathbb{C})}P_{mn}\otimes P_{ij}
\end{equation}
\noindent  is positive or not. To that end, let us define an element $Q$ of $\mathcal{C}_{2}(\mathfrak{H})$ as 

\begin{equation}
Q=\sum_{ij=1}^{\infty}\sum_{mn=1}^{\infty}q_{ijmn}P_{im}\otimes P_{jn}
\end{equation}

\noindent where the Fourier coefficients $q_{ijmn}$ are chosen so that

\begin{equation}
f_{nj}=\sum_{ab=1}^{\infty}q^{*}_{abnj}e_{ab}
\end{equation}

\noindent where $\{e_{ab}\}$ denotes an arbitrary Hilbert basis of $\ell^{2}(\mathbb{C})$. Then it is matter of simple computation to check that we can write 

\begin{equation}
\sum_{ij=1}^{\infty}\sum_{mn=1}^{\infty}(f_{nj},f_{mi})_{\ell^{2}(\mathbb{C})}P_{mn}\otimes P_{ij}=Q^{\dagger}Q
\end{equation}

\noindent Hence it is positive.\\

$(\Leftarrow)$ Let us now prove the converse, i.e.\ that if $\mathcal{J}_{e}[\alpha]$ is positive, then $\alpha$ is CP.\\

If $\mathcal{J}_{e}[\alpha]$ is positive, then there exists an element $Q\in\mathcal{C}_{2}(\mathfrak{H}_{2})$ such that 

\begin{equation}
\sum_{ij=1}^{\infty}\alpha[P_{ij}]\otimes P_{ij}=Q^{\dagger}Q
\end{equation}

This operator $Q$ can be expanded as 

\begin{equation}
Q=\sum_{ij=1}^{\infty}\sum_{mn=1}^{\infty}q_{ijmn}P_{im}\otimes P_{jn}
\end{equation}

Then define square-summable sequences $f_{nj}\in\ell^{2}(\mathbb{C})$ such that $f_{nj}=\sum_{ab=1}^{\infty}q^{*}_{abnj}e_{ab}$, where $\{e_{ab}\}$ denotes an arbitrary Hilbert basis of $\ell^{2}(\mathbb{C})$. Then performing the preceding computation in the reverse order, one arrives at 

\begin{equation}
\mathcal{J}_{e}[\alpha]=\sum_{ij=1}^{\infty}\sum_{mn=1}^{\infty}(f_{nj},f_{mi})_{\ell^{2}(\mathbb{C})}P_{mn}\otimes P_{ij}
\end{equation}

Everything is reduced to find a set of Hilbert-Schmidt Kraus operators $M_{p}$ such that 
\begin{equation}
\alpha[X]=\sum_{p=1}^{\infty}M_{p}XM_{p}^{\dagger}\quad\forall X\in\mathcal{C}_{2}(\mathfrak{H})
\end{equation}

Note that this is equivalent to know their Fourier coefficients $(P_{ij},M_{p})_{2}$ for all $i,j\geq 1$, which can be accomplished by defining $(P_{ij},M_{p})_{2}$, $p\geq 1$, $i,j$ fixed, as the entries of the sequence $f_{ji}$. Under this definition, we have 

\begin{equation}
(f_{nj},f_{mi})_{\ell^{2}(\mathbb{C})}=\sum_{p=1}^{\infty}(e_{n},M_{p}e_{j})_{\mathfrak{H}}^{*}(e_{m},M_{p}e_{i})_{\mathfrak{H}}
\end{equation}

\noindent which, using standard Hilbert space techniques, drives us to 

\begin{eqnarray}
\mathcal{J}_{e}[\alpha]&=&\sum_{ij=1}^{\infty}\sum_{mn=1}^{\infty}\left(P_{mn},\sum_{p=1}^{\infty}M_{p}P_{ij}M_{p}^{\dagger}\right)_{2}P_{mn}\otimes P_{ij}\nonumber\\
&=&\sum_{ij=1}^{\infty}\left[\sum_{p=1}^{\infty}M_{p}P_{ij}M_{p}^{\dagger}\right]\otimes P_{ij}
\end{eqnarray}

Thus $\alpha[P_{ij}]=\sum_{p=1}^{\infty}M_{p}P_{ij}M_{p}^{\dagger}$ for all $i,j\geq 1$, i.e.\ $\alpha[X]=\sum_{p=1}^{\infty}M_{p}XM_{p}^{\dagger}$ for all $X\in\mathcal{C}_{2}(\mathfrak{H})$. Hence the complete positivity of $\alpha$.

\end{proof}

The corollaries following this theorem in the finite-dimensional case are also valid in the infinite-dimensional case with the corresponding changes:

\begin{corol}\label{Cor1}
Let $\alpha$ be a CP map. Then $\mathcal{J}_{e}[\alpha]=Q^{\dagger}Q$, where $$Q=\sum_{ij=1}^{\infty}\sum_{mn=1}^{\infty}q_{ijmn}P_{im}\otimes P_{jn}$$ and the Fourier coefficients of $M_{p}$ in the Hilbert basis $P_{ij}\equiv|e_{i}\rangle\langle e_{j}|$ are given by the $p$th components of the square-summable sequences

\begin{equation}
f_{ij}=\sum_{ab=1}^{\infty}q^{*}_{abij}e_{ab}
\end{equation}

\noindent where $\{e_{ab}\}$ denotes an \emph{arbitrary} Hilbert basis of $\ell^{2}(\mathbb{C})$.
\end{corol} 

Note that this corollary entails the possibility of finding any possible set of Kraus operators, just by choosing another equally valid Hilbert basis $\{\tilde{e}_{ij}\}\equiv\{Ue_{ij}\}$ ($U$ a unitary operator).

\begin{corol}
The number of operators in a minimal Kraus representation of a CP linear map equals the number of strictly positive eigenvalues of $\mathcal{J}_{e}$.
\end{corol}

\begin{proof}
In case of an infinite minimal set of Kraus operators, it is clear that the point spectrum of $\mathcal{J}_{e}[\alpha]$ is countably infinite. Thus the only point of discussion arises when there is a finite minimal set of Kraus operators. But this case is also elementary, since under that hypothesis the Fourier expansion of $Q$ is truncated and turns into a sum. Thus, using the preceding corollary, it is immediate to realize that there will as many Kraus operators as non-zero eigenvalues of $\mathcal{J}_{e}[\alpha]$.
\end{proof}

\section{Examples}
\label{Exa}
We will show three elementary, though illustrative, examples. Firstly we will focus upon the transposition map\footnote{I.e.\ the composition of both complex and Hermitian conjugations.}  $\alpha[X]=X^{T}$. For this linear map, we have 

\begin{equation}
\mathcal{J}_{e}[\alpha]=\sum_{ij=1}^{\infty}P_{ij}\otimes P_{ji}
\end{equation}

\noindent which is non-positive. To prove this, just consider one of its eigenvectors, namely  $e_{p}\otimes e_{q}-e_{q}\otimes e_{p}$ which has a negative eigenvalue: $-1$. Thus the transposition is not CP, as it was already known by other means \cite{Arv69a}.\\

As a second example consider the map $\alpha[X]=\textrm{tr}\left(X\right)W$ on $\mathcal{C}_{1}(\mathfrak{H})$, where $W$ is a fixed positive unit-trace trace-class operator. For this linear map, we have

\begin{equation}
\mathcal{J}_{e}[\alpha]=W\otimes\mathbb{I}_{\mathfrak{H}}
\end{equation}

\noindent  which is clearly positive, thus $\alpha$ is CP. Using the spectral decomposition of $W=\sum_{n=1}^{\infty}w_{nn}P_{nn}$ in a given Hilbert orthonormal basis $\{e_{i}\}$ of $\mathfrak{H}$, we can write $\mathcal{J}_{e}[\alpha]=Q^{\dagger}Q$ with 

\begin{equation}
Q=\sum_{ij=1}^{\infty}\sum_{mn=1}^{\infty}q_{ijmn}P_{im}\otimes P_{jn}=\sum_{n=1}^{\infty}w_{nn}^{1/2}P_{nn}\otimes\mathbb{I}_{\mathfrak{H}}
\end{equation}

\noindent i.e.\ $q_{ijmn}=w_{ii}^{1/2}\delta_{im}\delta_{jn}$ and the square-summable sequences of corollary \ref{Cor1} are then given by

\begin{equation}
f_{ij}=\sum_{ab=1}^{\infty}w_{aa}^{1/2}\delta_{ai}\delta_{bj}e_{ab}=w_{ii}^{1/2}e_{ij}
\end{equation}

\noindent Under these conditions, choosing as $e_{ij}$ the canonical basis $\{e_{k}\}$ of $\ell^{2}(\mathbb{C})$ so that $P_{ij}=|e_{i}\rangle\langle e_{j}|$, a valid set of Kraus operators is given by

\begin{equation}
M_{ij}=w_{ii}^{1/2}P_{ij}
\end{equation}

\noindent so that $\alpha[X]=\sum_{ij=1}^{\infty}M_{ij}XM_{ij}^{\dagger}$. An alternative set can be obtained by choosing another Hilbert basis of $\ell^{2}(\mathbb{C})$, i.e.\ $\{Ue_{k}\}$, where $U$ denotes a unitary operator.\\

As a final example consider the map $\alpha[X]=\mu\textrm{tr}\left(X\right) W+(1-\mu)X$ on the space of trace-class operators and where $W$ is a fixed positive unit-trace trace-class operator and $\mu\in[0,1]$. When applied to the convex cone of selfadjoint unit-trace positive operators on $\mathfrak{H}$ (i.e.\ the set of density operators of a quantum system with associated Hilbert space $\mathfrak{H}$ \cite{GalPas89a}), this map reduces to $\alpha[\rho]=\mu W+(1-\mu)\rho$. This is a generalization of the depolarizing channel used in quantum information theory \cite{NieChu00a}. Its CP character arises from the convex structure of the set of CP maps\footnote{Which can be proven using the Kraus decomposition.} \cite{Cho72a} and the complete positivity of the previous map and the identity map. 

\section{Discussion and Conclusions}
\label{DisCon}
As a first comment, it should be noticed that the existence of a Jamio\l kowski criterion for linear maps between algebras of Hilbert-Schmidt operators is sufficient for the study of complete positivity in Quantum Physics. All one has to realize is that the set of quantum states is a convex cone of the space of trace-class operators $\mathcal{C}_{1}(\mathfrak{H})$, which is contained in the space of Hilbert-Schmidt class operators $\mathfrak{C}_{2}(\mathfrak{H})$ (Theorem 9.19 of \cite{AbeGal91a}). From the strictly mathematical point of view it could be desirable to have a proof for arbitrary $C^{*}$-algebras of operators, so that this question still remains open.\\

As briefly commented in the introduction, two are the main applications of these tools, namely (i) the analysis of entanglement and separability of quantum systems, which, under this criterion, is translated to the study of positive maps between algebras of operators. This is surprisingly still an open question in linear algebra, where only a generic decomposition for the particular cases $\mathbb{C}^{2}\otimes\mathbb{C}^{2}$ and $\mathbb{C}^{2}\otimes\mathbb{C}^{3}$ has been found \cite{Wor76a}, and (ii) the dynamics of open quantum systems, in which the question of the complete positivity of an arbitrary reduced dynamics also remains to be definitively answered.

\section*{Acknowledgments}
This work is dedicated to Prof.\ A.\ Galindo, a pioneer of the research in Quantum Physics in Spain, on the occasion of his 70th birthday. We acknowledge financial support of Spanish Ministry of Science and Technology under project No.\ FIS2004-01576. %M.F. also acknowledges support from Oviedo University (ref.\ no.\ MB-04-514).

%\subsection{Subsection Heading}
%\label{sec:2}
%Your text goes here.

%\begin{equation}
%\vec{a}\times\vec{b}=\vec{c}
%\end{equation}

%\subsubsection{Subsubsection Heading}
%Your text goes here. Use the \LaTeX\ automatism for cross-references as
%well as for your citations, see Sect.~\ref{sec:1}.

%\paragraph{Paragraph Heading} %
%Your text goes here.

%\subparagraph{Subparagraph Heading.} Your text goes here.%
%\begin{table}
%\centering
%\begin{tabular}{lll}
%\hline\noalign{\smallskip}
%\end{tabular}
%\end{table}
%\begin{figure}
%\centering
%\includegraphics[height=4cm]{figure.eps}
%\picplace{5cm}{2cm} % Give the correct figure height and width in cm
%\end{figure}
%
% For built-in environments use
%

\end{document}